\documentclass[lettersize,journal]{IEEEtran}
\usepackage{amsmath,amsfonts}
\usepackage{algorithmic}
\usepackage{algorithm}
\usepackage{array}
\usepackage{bm}
\usepackage{stfloats}

\ifCLASSOPTIONcompsoc
  \usepackage[caption=false,font=normalsize,labelfont=sf,textfont=sf]{subfig}
\else
  \usepackage[caption=false,font=footnotesize]{subfig}
\fi
\usepackage{multirow}
\usepackage{booktabs}
\usepackage{textcomp}
\usepackage{stfloats}
\usepackage{url}
\usepackage{verbatim}
\usepackage{graphicx}
\usepackage{subcaption}
\usepackage{cite}
\usepackage{color} 
\usepackage{hyperref}
\usepackage{amssymb}
\usepackage{rotating}
\usepackage[detect-all]{siunitx}
\usepackage{etoolbox}
\usepackage[table]{xcolor}

\DeclareMathOperator*{\argmin}{arg\,min}

\begin{document}
\title{Structural Similarity-Inspired Unfolding for Lightweight Image Super-Resolution}

\author{
    Zhangkai Ni,~\IEEEmembership{Member,~IEEE}, 
    Yang Zhang,
    Wenhan Yang,~\IEEEmembership{Member,~IEEE}, \\
    Hanli Wang,~\IEEEmembership{Senior Member,~IEEE}, 
    Shiqi Wang,~\IEEEmembership{Senior Member,~IEEE}, 
    Sam Kwong,~\IEEEmembership{Fellow,~IEEE}

\thanks{This work was supported in part by the National Natural Science Foundation of China under Grant 62201387, and Grant 62371343, in part by the Fundamental Research Funds for the Central Universities, and in part by the Interdisciplinary Frontier Research Project of PCL under Grant 2025QYB013. 
\emph{(Corresponding authors: Hanli Wang and Wenhan Yang)}
}
\thanks{Zhangkai Ni, Yang Zhang, and Hanli Wang are with the School of Computer Science and Technology and the Key Laboratory of Embedded System and Service Computing (Ministry of Education), Tongji University, Shanghai 200092, China (e-mail: zkni@tongji.edu.cn; zhangy\_ce@tongji.edu.cn; hanliwang@tongji.edu.cn).}
\thanks{Shiqi Wang is with the Department of Computer Science, City University of Hong Kong, Hong Kong 999077 (e-mail: shiqwang@cityu.edu.hk).}
\thanks{Wenhan Yang is with Pengcheng Laboratory, Shenzhen, Guangdong 518066, China. (e-mail: yangwh@pcl.ac.cn).}
\thanks{Sam Kwong is with the School of Data Science, Lingnan University, Hong Kong (e-mail: samkwong@ln.edu.hk).}
}

\markboth{Journal of \LaTeX\ Class Files,~Vol.~14, No.~8, August~2021}%
{Shell \MakeLowercase{\textit{et al.}}: A Sample Article Using IEEEtran.cls for IEEE Journals}

\maketitle

\begin{abstract}
Major efforts in data-driven image super-resolution (SR) primarily focus on expanding the receptive field of the model to better capture contextual information.
However, these methods are typically implemented by stacking deeper networks or leveraging transformer-based attention mechanisms, which consequently increases model complexity.
In contrast, model-driven methods based on the unfolding paradigm show promise in improving performance while effectively maintaining model compactness through sophisticated module design.
Based on these insights, we propose a Structural Similarity-Inspired Unfolding (SSIU) method for efficient image SR. This method is designed through unfolding an SR  optimization function constrained by structural similarity, aiming to combine the strengths of both data-driven and model-driven approaches.
Our model operates progressively following the unfolding paradigm. Each iteration consists of multiple Mixed-Scale Gating Modules (MSGM) and an Efficient Sparse Attention Module (ESAM). The former implements comprehensive constraints on features, including a structural similarity constraint, while the latter aims to achieve sparse activation.
In addition, we design a Mixture-of-Experts-based Feature Selector (MoE-FS) that fully utilizes multi-level feature information by combining features from different steps.
Extensive experiments validate the efficacy and efficiency of our unfolding-inspired network.
Our model outperforms current state-of-the-art models, boasting lower parameter counts and reduced memory consumption. 
Our code will be available at: \url{https://github.com/eezkni/SSIU}
\end{abstract}


\section{Introduction}
\label{sec1}
\IEEEPARstart{I}{MAGE} Super-Resolution (SR) aims to reconstruct High-Resolution (HR) images from Low-Resolution (LR) observations by generating sharp, clear edges and revealing fine texture details while suppressing visual artifacts~\cite{srcnn}.
It has a wide range of applications, such as improving the image quality to provide a better human visual experience for medical imaging, security inspection, and image compression, or facilitating downstream computer vision tasks such as object detection and recognition.
In recent years, deep learning-based SR methods have gained attention, leveraging large-scale data to obtain the mapping between LR and HR images. 
One of the most important strategies employed by these methods is the expansion of the network's receptive field.
Expanding the receptive field of the network allows it to cover a larger range of input information, significantly improving its capacity to capture long-range dependencies and structural information in the data.
Common strategies for enlarging the receptive field include increasing the number of convolutional layers~\cite{vdsr,  cai2022tdpn}, 
though this can lead to challenges such as vanishing gradients, which calls for the design of additional modules.
An alternative approach adopts attention mechanisms~\cite{rcan, san, han, huang2021interpretable,xu2024fdsr}, 
allowing the model to focus on relevant image features and improve SR performance.
Transformer-based SR methods~\cite{ni2024m2trans, esrt, swinir, cai2023hipa} then emerged, effectively leveraging self-attention to capture global information, thus enhancing their ability to perceive contextual cues.  
These methods expand receptive fields to activate more neighboring pixels, resulting in substantial performance improvements. 
However, with increasingly complex module designs, these methods often lead to greater network depth and heightened computational and memory demands.

Lightweight SR networks are crucial in mitigating model resource consumption, offering pathways for efficient computation. 
Various methods have been developed to address specific challenges in this area.
Some methods focus on designing lightweight modules~\cite{subpixel, zhang2020adaptive}, 
while others apply network compression~\cite{xu2023spl} or distillation~\cite{zhang2021cvpr} to derive low-complex models from existing architectures.
Another key class of methods enhances training constraints by incorporating domain knowledge, thereby improving the efficiency of module design and integration to reduce the complexity of the model.
Notable examples include crafting trainable unfolding networks, as studied in works such as \cite{zhang2022tpami, zhang2019cvpr, zhou2023memory}. 
These methods combine model-driven domain insights with learning-based paradigms via decomposing complex inverse problems into sequential subtasks, each solved by a dedicated neural network, resulting in end-to-end SR models with enhanced efficiency and performance.
From module design optimization to domain knowledge integration, these approaches collectively advance the field, offering efficient and interpretable solutions tailored to lightweight image SR.

From the above reviews, we derive two key insights:
1) Expanding the receptive field is a double-edged sword. While it effectively enhances SR performance, it also significantly increases model complexity.
2) In lightweight SR research, the unfolding paradigm-based approach offers a more general, practical, and interpretable solution.
Inspired by these observations, our work introduces the unfolding paradigm to decompose the image SR problem to construct a progressive, trainable, and efficient deep neural network. 
Specifically, the Maximum A Posteriori (MAP) based image SR problem can be expressed as follows:
\begin{align}
\label{eq:degradation_intro}
\mathbf{y} &=\mathbf{H} \mathbf{x}+ \mathbf{n}, \\
\hat{\mathbf{x}} & = \argmin_{ \mathbf{\mathbf{x}}} \left\{ 
\| \mathbf{y}-\mathbf{H} \textbf{x} \|_2^2 + \lambda p\left( \textbf{x} \right) \right\},
\label{eq:degradation_sr}
\end{align}
where $\mathbf{x}$ is the HR image, $\mathbf{y}$ is the degraded LR image, $\mathbf{n}$ represents noise, $\mathbf{H}$ is a degradation matrix, $p (\cdot)$ is visual prior regularization term, and $\hat{\mathbf{x}}$ is the obtained solution.
The Equ.~\eqref{eq:degradation_intro} formulates the degradation process, and Equ.~\eqref{eq:degradation_sr} is the SR estimation process.
In Equ.~\eqref{eq:degradation_sr}, 
from practical experience, $\mathbf{H}$ is observed to be local and dynamic, while $p(\cdot)$ usually presents global context constraint.
Several studies have considered solving the inverse problem with these two properties in a unified way.  
For instance, HAT~\cite{hat} combines local and global information to activate more distant pixels, leading to improved SR results.
Traditional methods based on self-similarity and nonlocal similarity~\cite{dong2011image} suggest that only structurally similar distant information benefits the SR performance of central pixels.
Therefore, we propose a Structural Similarity-Inspired Unfolding network to focus on regions with similar structures to the central pixel with global information aggregation.

\begin{figure}[t]
    \centering
    \centerline{\includegraphics[width=1.0\linewidth]{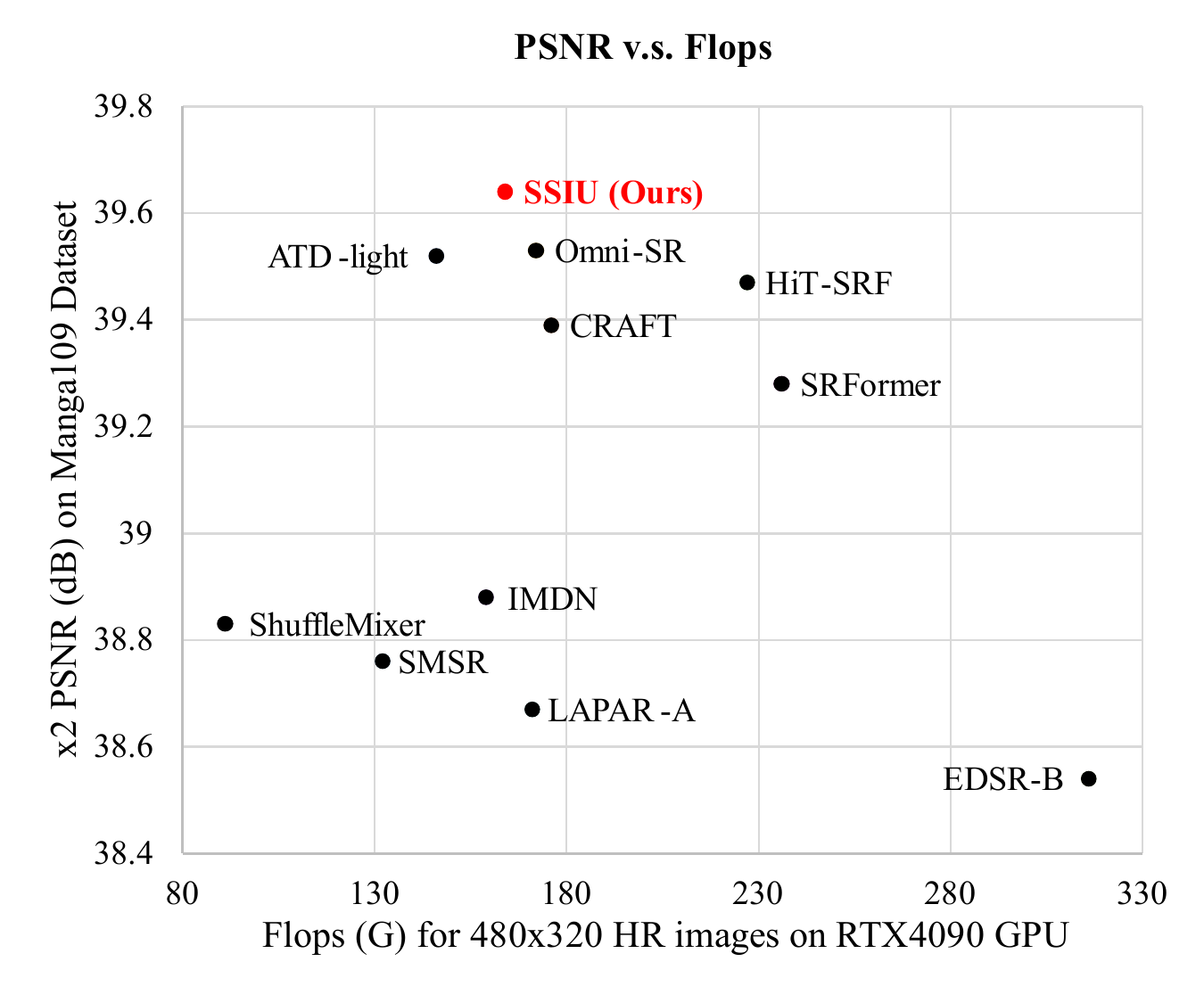}}
    \caption{Performance comparison on the Manga109 dataset. The proposed SSIU model achieves a favorable balanced trade-off between PSNR and FLOPs.
    }
    \label{fig:cmp}
\end{figure}

Our method achieves a favorable balance between reconstruction performance and computational efficiency by combining the strengths of both convolutional and transformer-based methods, along with data-driven and model-driven approaches, as shown in Fig.~\ref{fig:cmp}.
Specifically, our network is derived from a typical unfolding-based non-local centralized sparse coding problem. 
In the progressive solution process, we design a mixed-scale gating module to impose constraints on feature representation and a sparse attention mechanism to achieve efficient sparse activation. 
Additionally, we introduce a mixture-of-experts-based feature selector for gating features from different scales.
This progressive framework and the three proposed modules efficiently integrate global and local information, improving SR performance while maintaining a lightweight architecture.
In summary, the key contributions are as follows:
\begin{itemize}
    \item  
    We propose a novel Structural Similarity-Inspired Unfolding (SSIU) network for lightweight image SR, which sparsely activates long-range pixels, leading to improved SR results with a lightweight architecture.
    
    \item 
    We design a Mixed-Scale Gating Module (MSGM) to impose sparsity and structural similarity constraints, and an Efficient Sparse Attention Module (ESAM) to reduce the complexity of long-range modeling.
    
    \item 
    We introduce a Mixture-of-Experts Feature Selector (MoE-FS) that effectively integrates multi-scale features using a learnable gating mechanism. Extensive experiments show that SSIU outperforms state-of-the-art models, achieving better SR performance with faster inference and lower memory usage.
\end{itemize}

The rest of this paper is organized as follows. Section~\ref{sec2} provides a concise review of related work. Section~\ref{sec4:proposed} details the proposed SSIU model.  Section~\ref{sec5:results} presents the experimental results. Section~\ref{sec:conclusion} concludes the paper.

\section{Related works}
\label{sec2}
\subsection{Deep Networks for Image Super-Resolution}
With the advent of deep learning, CNN-based SR models~\cite{srcnn,vdsr,drrn,huang2021interpretable,cai2022tdpn,xu2024fdsr,yan2024kgsr} have become the prevailing approach. 
Dong et al. introduced SRCNN~\cite{srcnn}, the first CNN-based SR network, which outperformed traditional methods and laid the groundwork for deep SR. 
VDSR~\cite{vdsr} then employed a very deep network to achieve high-precision reconstruction. 
DeFiAN~\cite{huang2021interpretable} incorporated a fidelity attention mechanism to adaptively enhance both high-frequency details and low-frequency smoothness. 
TDPN~\cite{cai2022tdpn} added a texture-guided branch to improve perceptual quality, and FDSR~\cite{xu2024fdsr} operates in the frequency domain by dividing images into distinct frequency bands for targeted reconstruction. 
Although CNN-based methods have advantages in terms of computational complexity, their performance is insufficient.

Based on Vision Transformer (ViT)~\cite{vit}, several methods~\cite{ttsr,swinir,esrt,hat,act,zhang2024transcending} have explored Transformer-based SR. TTSR~\cite{ttsr} proposed the first Transformer-based SR model, using hard and soft attention modules for reference-based SR. SwinIR~\cite{swinir} advanced SR by combining the Swin Transformer's local window self-attention with convolutional operations.
ESRT~\cite{esrt} optimized the structure of the Transformer by fusing CNN and Transformer features to reduce complexity and improve performance. HAT~\cite{hat} improved performance by integrating channel attention with self-attention, introducing an overlapping cross-attention module and a pre-training strategy. 
ACT~\cite{act} proposed a parallel CNN-Transformer architecture based on cross-scale attention for extracting both local and global features.
ATD~\cite{zhang2024transcending} introduced an adaptive token dictionary to recover lost high-quality details.
Despite these advancements, Transformer-based models still demand considerable resources due to the computational cost of the self-attention mechanism.
Inspired by recent image enhancement studies~\cite{yan2024dynamic,yan2025efficient}, several works have explored the integration of diffusion models into SR~\cite{niu2024acdmsr,wang2024sinsr}.
ACDMSR~\cite{niu2024acdmsr} introduced a diffusion-based SR framework that leverages a pre-trained SR model to provide improved conditional inputs.
SinSR~\cite{wang2024sinsr} accelerated the diffusion-based SR model to a single inference step based on a deterministic sampling process.

\subsection{Lightweight Networks for Image Super-Resolution}
To reduce model resource consumption, various CNN-based methods have been developed~\cite{edsr,carn,imdn,lapar,ecbsr,smsr,shufflemixer}.
These include EDSR~\cite{edsr}, which simplifies residual networks by removing unnecessary modules; CARN~\cite{carn}, which adopts a cascaded residual structure; and IMDN~\cite{imdn}, which utilizes multi-distillation for faster inference.
LAPAR~\cite{lapar} formulates SR as a linear regression problem for efficient enhancement, while ECBSR~\cite{ecbsr} focuses on edge-aware lightweight design for mobile devices.
SMSR~\cite{smsr} leverages sparse masking for computational efficiency, and ShuffleMixer~\cite{shufflemixer} explores large-kernel convolutions and channel splitting.
However, CNN-based methods are inherently limited by local receptive fields, which constrain their capacity for long-range modeling.

Motivated by the strong capability of Transformers in capturing global dependencies, a series of recent works~\cite{lbnet, zhou2023srformer, li2023feature, SAFMN, wang2023omni, zhang2024hitsr} have investigated their use in enhancing image SR performance.
LBNet~\cite{lbnet} combines a symmetric CNN and a recursive Transformer to extract local features and long-range dependencies, effectively balancing performance and computational cost.  
SRFormer~\cite{zhou2023srformer} proposes a permuted self-attention mechanism that shifts spatial information into the channel dimension, reducing the overhead introduced by large attention windows.  
CRAFT~\cite{li2023feature} incorporates high-frequency priors to guide SR, thereby simplifying network design.  
SAFMN~\cite{SAFMN} replaces multi-head attention with a multi-scale feature modulation module to reduce computational burden.  
Omni-SR~\cite{wang2023omni} adopts a full self-attention scheme to model spatial and channel interactions jointly, achieving both lightweight design and improved performance.  
HiT-SR~\cite{zhang2024hitsr} introduces a hierarchical Transformer to capture multi-scale features and long-range dependencies, further boosting SR quality.  
Although these methods reduce complexity through architectural innovations, they primarily concentrate on network design rather than explicitly modeling the SR reconstruction process to achieve further efficiency improvements.

\section{Structural Similarity-Inspired Unfolding Network}
\label{sec4:proposed}

\subsection{Problem Formulation}
\label{sec4_1:unfolding}
We formulate the image SR problem as a MAP estimation problem with structural similarity constraints. 
Given an LR image $\mathbf{y}$ and an HR image $\mathbf{x}$, the image SR process can be expressed using MAP estimation, as shown in Equ.~\eqref{eq:degradation_sr}. 
Following the setting of the Sparse Coding (SC) model ~\cite{dong2011image}, we simplify the MAP estimation problem into a sparse coding problem based on the $L_1$-norm:
\begin{equation}
    \alpha_{\mathbf{x}} = \argmin_{ \mathbf{\alpha}} \left\{  \| \mathbf{y}-\mathbf{H}\mathbf{\Phi} \mathbf{\alpha} \|_2^2 + \lambda \|\mathbf{\alpha}\|_1 \right\},  
\end{equation}
where $\mathbf{\Phi}$ is a signal dictionary with signal atoms that help reconstruct the HR images.
$\mathbf{\alpha}$ is the sparse coding coefficient representing the weight of dictionary atoms, $\mathbf{\alpha}_{\mathbf{x}}$ is the predicted sparse coding coefficient, and $\lambda$ is the weighting parameter.
The predicted HR image $\mathbf{\hat{x}}$ can be obtained using $\mathbf{\alpha}_{\mathbf{x}}$ and $\Phi$: $\hat{\mathbf{x}} = \Phi \mathbf{\alpha}_{\mathbf{x}}$.
Inspired by the traditional methods of using non-local similarity~\cite{NCSR}, we further impose structural similarity constraints on the coefficient $\alpha$ to pay more attention to the information of structurally similar regions:
\begin{align}
    \alpha_{\mathbf{x}}  = \argmin_{\alpha} \left\{ \| \mathbf{y}-\mathbf{H}\mathbf{\Phi} \mathbf{\alpha} \|_2^2 + \lambda \| \mathbf{\alpha} \|_1  + \gamma  \|\mathbf{\alpha} - \mathbf{\beta} \|_1 \right\}, 
\end{align}
where the coefficient $\beta$ is a variable estimated from $\alpha$ based on structural similarity, and $\gamma$ is a weighting parameter. 
To facilitate optimization, we introduce an auxiliary variable $\mathbf{z}$ to replace $\alpha$, resulting in the following optimization problem:
\begin{align}
\label{whole}
    \alpha_{\mathbf{x}} = & \argmin_{\alpha} \left\{ \| \mathbf{y}-\mathbf{H}\mathbf{\Phi} \mathbf{\alpha} \|_2^2 + \lambda  \| \mathbf{z} \|_1 \nonumber + \gamma \| \mathbf{\alpha} - \mathbf{\beta} \|_1 \right.\\
    & \left. + \eta \|\mathbf{z} - \mathbf{\alpha} \|_2 \right\},
\end{align}
where $\eta$ is a weighting parameter.

\begin{figure*}[t]
    \centering
    \centerline{\includegraphics[width=1.0\linewidth]{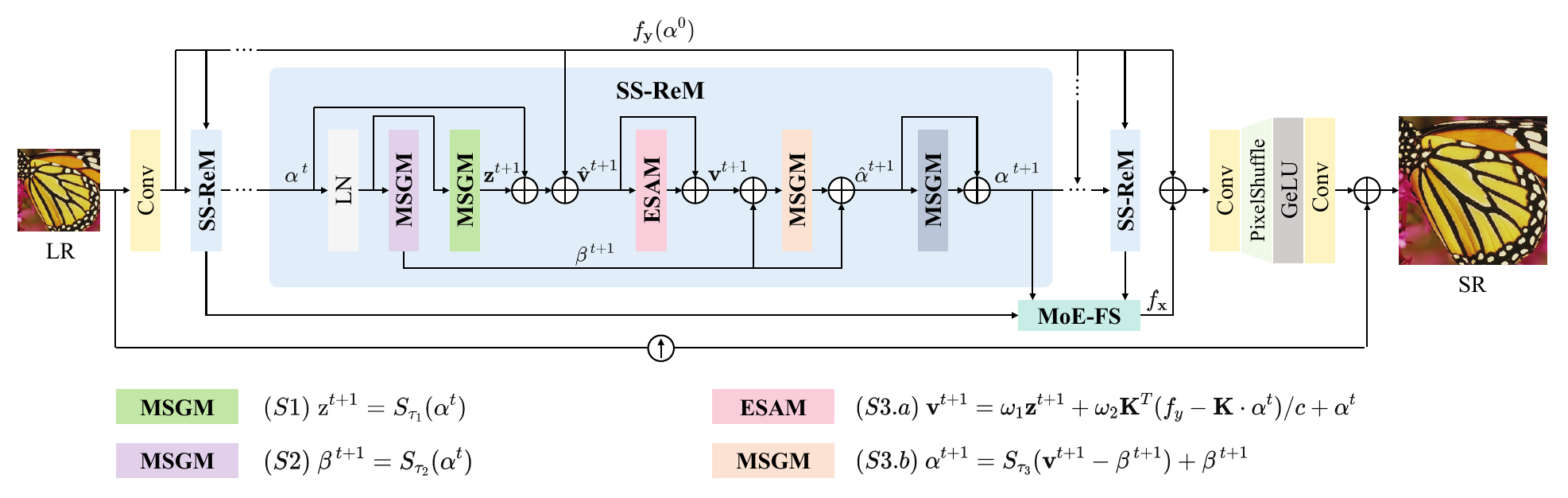}}
    \caption{
    Overall pipeline of the SSIU. 
    It progressively refines features using the Structural Similarity-Inspired Recurrent Module (SS-ReM), which integrates Mixed-Scale Gating Modules (MSGM) for sparsity and structural similarity constraints, and the Efficient Sparse Attention Module (ESAM) for efficient long-range modeling. 
    The Mixture-of-Experts-based Feature Selector (MoE-FS) further refines feature by gating outputs from multiple SS-ReM modules. 
    For simplicity, we retain the symbols $\alpha$, $\beta$, $\mathbf{v}$, and $\mathbf{z}$ to represent their corresponding features, while converting only the input $\mathbf{y}$ into the feature representation $f_{\mathbf{y}}$.
    }
    \label{fig:framework}
\end{figure*}

To solve the optimization problem in Equ.~\eqref{whole}, we employ an alternating optimization approach where the value of one variable is updated while keeping the other variables fixed. 
This approach partitions the problem into three univariate subproblems, which can be optimized using the following alternating scheme at stage $t+1$:
\begin{align}
    &(P1)~\mathbf{z}^{t+1} =  \argmin_{\mathbf{z}} 
	\left\{ \lambda \| \mathbf{z} \|_1  +  \eta \| \mathbf{z} - \mathbf{\alpha}^t\|_2 \right\}, \nonumber \\
    &(P2)~\beta^{t+1} =\argmin_{\mathbf{\beta}} \left\{ \gamma \| \mathbf{\alpha}^t - \mathbf{\beta} \|_1 \right\}, \nonumber\\
    &(P3)~\alpha^{t+1} =  \argmin_{\alpha} \left\{ \| \mathbf{y} -\mathbf{H}\mathbf{\Phi}\alpha \|_2^2 + \gamma \|\alpha - \beta^{t+1} \|_1 \right.\nonumber\\
    &~\quad \quad + \left. \eta \| \mathbf{z}^{t+1} - \alpha\|_2 \right\}. \nonumber
\end{align}

Thus, an iterative solution can be obtained to update $\mathbf{z}^{t}$, $\mathbf{\beta}^{t}$ and $\mathbf{\alpha}^{t}$ sequentially with Half Quadratic Splitting (HQS)~\cite{he2013half}:
\begin{align}
    &(S1)~\mathbf{z}^{t+1} = \rm S_{\tau_1}\left( \mathbf{\alpha}^t \right),  \nonumber \\
    &(S2)~\mathbf{\beta}^{t+1} = \rm S_{\tau_2}\left( \mathbf{\alpha}^t \right),  \nonumber \\
    &(S3.a)~\mathbf{v}^{t+1} = \omega_1 \mathbf{z}^{t+1} + \omega_2 \mathbf{K}^{\rm T}\left( \mathbf{y} - \mathbf{K} \cdot \alpha^t  \right)/c + \alpha^t, \nonumber \\
    &(S3.b)~\alpha^{t+1} = \rm S_{\tau_3} \left( \mathbf{v}^{t+1} - \beta^{t+1} \right) + \beta^{t+1}, \nonumber 
\end{align}
where $\rm S_{\tau}( \cdot )$ is the soft threshold operator, representing the solution process. $\tau_1$, $\tau_2$, and $\tau_3$ are hyperparameters related to the constraints.
$\mathbf{v}$ is an intermediate variable to facilitate the description of subproblem $(S3)$.
$\mathbf{K} = \mathbf{H} \mathbf{\Phi}$, $\mathbf{K}^{\rm T} = \mathbf{\Phi}^{\rm T} \mathbf{H}^{\rm T}$, ${\rm T}$ represents the transpose operation.
$\omega_1$ and $\omega_2$ are weighting parameters and $c$ is an auxiliary parameter guaranteeing the convexity of the surrogate function~\cite{daubechies2004iterative}.
In summary, $(S1)$ applies a sparse constraint on $\alpha$ and $(S2)$ imposes structural similarity constraint. 
Then $(S3.a)$ performs a global sparse inverse-solving process and $(S3.b)$ centralizes the estimation to $\mathbf{\alpha}^{t+1}$.
Finally, after $N$ iterations, we have $\alpha_{\mathbf{x}} = \alpha^{N}$, and the corresponding SR result is $\hat{\mathbf{x}} = \mathbf{\Phi} \alpha_{\mathbf{x}}$.

Inspired by the above solution, we follow the general route of the processes ($S1$, $S2$, $S3$) to construct our network.
It is worth noting that the solution of the above unfolding paradigm is performed in the image space, and the LR image $\mathbf{y}$ participates in the solution process. In this paper, we use the unfolding paradigm in the feature space. The feature space retains the key structural information of the image space, ensuring that the above solution still holds.
To simplify the symbolic representation, we continue to use $\alpha$, $\beta$, $\mathbf{v}$, $\mathbf{z}$ to represent the corresponding features and only convert the input $\mathbf{y}$ into the feature $f_{\mathbf{y}}$.
As illustrated in Fig.~\ref{fig:framework}, our proposed model performs the following process at each stage $t+1$:
\begin{itemize}
    \item \textbf{S1}: We impose a sparse constraint on the feature $\alpha^t$ from the previous stage to update $\mathbf{z}^{t}$. 
    This operation is simulated using a mixed-scale gating module (MSGM).
    \item \textbf{S2}: The feature $\alpha^{t}$ is fed into another MSGM to obtain the estimate $\beta^{t+1}$ of $\alpha^{t}$ based on structural similarity.
    \item \textbf{S3.a}: The feature $\alpha^t$, the feature $\mathbf{z}^{t+1}$ from S1, and the feature $f_y$ extracted from $\mathbf{y}$ are processed and fused through an efficient sparse self-attention module (ESAM).    
    \item \textbf{S3.b}: Aggregate the constraints on feature $\textbf{v}^{t+1}$ and feature $\beta^{t+1}$ onto feature $\alpha^{t+1}$ through MSGM.
\end{itemize}

It is noted that we use the same module MSGM to solve subproblems S1, S2, and S3.b as the essence of these three subproblems lies in performing a linear transformation on the input followed by thresholding. 
Our MSGM is implemented based on a gating mechanism, which can achieve all targets of sparsity constraints, structural similarity constraints, and feature aggregation by dynamically adjusting the weights of input features.

\subsection{Overview of SSIU}
The overall pipeline of SSIU is illustrated in Fig.~\ref{fig:framework}, consisting of three key components: \textbf{Shallow Feature Extraction}, \textbf{Progressive Feature Refinement}, and \textbf{HR Image Reconstruction}, which are detailed as follows:

\subsubsection{Shallow Feature Extraction}
Given an LR image $\mathbf{y} \in R^{3\times h\times w}$ with spatial resolution $h\times w$, it is first embedded into feature $f_{\mathbf{y}}\in R^{C\times H\times W}$ through a shallow feature extraction module $\rm M_{FE}(\cdot)$. $C$ denotes the number of feature channels and $H\times W$ is the feature spatial resolution:
\begin{equation}
    f_{\mathbf{y}} = {\rm M_{FE} }(\mathbf{y}),
\end{equation}
where $\rm M_{FE} (\cdot)$ consists of a $3\times3$ convolution.

\subsubsection{Progressive Feature Refinement}
We perform progressive feature refinement using stacked Structural Similarity-Inspired Recurrent Modules (SS-ReM) to transform LR features into HR features. 
Each SS-ReM takes $\alpha$ and $f_{\mathbf{y}}$ as input and outputs the updated $\alpha$:
\begin{equation}
\alpha^{t+1} = {\rm \text{SS-ReM}_{t}}(\alpha^{t},f_{\mathbf{y}}), \quad t=0,1,...,N-1,
\end{equation}
where $\alpha^{t}$ is the output of the previous SS-ReM module, and $\alpha^{0}$ is initialized with the shallow features $f_{\mathbf{y}}$.
N means the number of SS-Rem and is set to 9 in this paper.
$\beta^{t}$, $\mathbf{v}^{t}$ and $\mathbf{z}^{t}$ are optimized by SS-ReM during the update process of $\alpha^{t}$.
In addition, rather than directly using the final output $\alpha^{N}$ from the last module, our method combines multi-level features, as their distinct characteristics lead to better enhancement results.
Therefore, we introduce a Mixture-of-Experts-based Feature Selector (MoE-FS), which learns to gate the features $\alpha^{t}$ from multiple stages to produce the final feature $f_{\mathbf{x}} \in R^{C\times H\times W}$:
\begin{equation}
f_{\mathbf{x}} = {\rm \text{MoE-FS}}(\alpha^{1},\cdots,\alpha^{N}).
\end{equation}

In order to balance the model performance and computational complexity, we only select 3 outputs of SS-ReM as the inputs of MoE-FS during implementation.

\subsubsection{HR Image Reconstruction}
The progressively refined features $f_\mathbf{x}$ are combined with the $f_\mathbf{y}$ to reconstruct the SR image.
Instead of learning the direct mapping from the LR image to the HR image, our model employs a reconstruction module ${\rm{M}_{RC}(\cdot)}$ to generate the residual image. 
The ${\rm{M}_{RC}(\cdot)}$ contains a $1 \times 1$ convolutional layer, a pixel shuffle layer, a GeLU activation layer, and a $3 \times 3$ convolutional layer. 
This process can be expressed as follows:
\begin{equation}
 \hat{\mathbf{x}}= {\rm{M}_{RC}}(f_{\mathbf{x}}+f_{\mathbf{y}}) + \Gamma(\mathbf{y}),
\end{equation}
where $\Gamma(\cdot)$ denotes bilinear upsampling, $\hat{\mathbf{x}} \in R^{3\times hs\times ws}$ is the final reconstructed image, and $s$ represents the SR scale.
The proposed SSIU achieves an excellent trade-off between computational overhead and reconstruction quality.

\subsection{Structural Similarity-Inspired Recurrent Module}
The stacked SS-ReMs are used to implement the progressive feature refinement process. 
As shown in Fig.~\ref{fig:framework}, each SS-ReM consists of several MSGM and an ESAM, corresponding to the steps $(S1, S2, S3)$.
Specifically, the feature $\alpha^{t}$ is processed by an MSGM after layer normalization (LN), imposing a sparsity constraint and yielding an updated auxiliary variable $\mathbf{z}^{t+1}$:
\begin{equation}
{\mathbf{z}}^{t+1} = {\rm S_{\tau_{1}}}(\alpha^{t}) = {\rm MSGM_{1}}(\rm LN (\alpha^{t})).
\end{equation}

Simultaneously, $\alpha^{t}$ is also input into another MSGM to obtain an estimate $\beta^{t+1}$ with structural similarity constraint:
\begin{equation}
\beta^{t+1} = {\rm S_{\tau_{2}}}(\alpha^{t}) = {\rm MSGM_{2}}(\rm LN (\alpha^{t})).
\end{equation}

Next, an ESAM is used to perform sparse activation on $\mathbf{z}^{t+1}$, $\alpha^{t}$, and $f_{\mathbf{y}}$ to obtain the updated intermediate variable $\mathbf{v}^{t+1}$. To simplify this module, we first sum the inputs into $\mathbf{\hat v}^{t+1}$ and then feed it into ESAM with a skip connection:
\begin{align}
\mathbf{\hat v}^{t+1} &= \mathbf{z}^{t+1} + \alpha^{t} + f_{\mathbf{y}},\nonumber \\
\mathbf{v}^{t+1} &= \rm ESAM(\mathbf{\hat v}^{t+1}) + \mathbf{\hat v}^{t+1}.
\end{align}
$\mathbf{v}^{t+1}$ is a combination of three features $\mathbf{z}^{t+1}$, $\alpha^{t}$, and $f_{\mathbf{y}}$, which is essentially the same as $S3.a$.
Subsequently, $\mathbf{v}^{t+1}$ and $\beta^{t+1}$ are input into an MSGM for feature aggregation to obtain feature $\hat{\alpha}^{t+1}$ with structural similarity constraint and sparse activation. Following $S3.b$, we input the sum of $\mathbf{v}^{t+1}$ and $\beta^{t+1}$ into MSGM and add the output to $\beta^{t+1}$:
\begin{align}
    \hat \alpha^{t+1} &= {\rm S}_{\tau_{3}}({\mathbf{v}}^{t+1} + \beta^{t+1}) + \beta^{t+1} \nonumber\\
    &= {\rm MSGM}_{3}({\mathbf{v}}^{t+1} + \beta^{t+1})+ \beta^{t+1}.
\end{align}

Finally, to further improve the model performance, $\hat{\alpha}^{t+1}$ is input into an MSGM-based Feedforward Network (FFN) for feature refinement:
\begin{equation}
     \alpha^{t+1} = {\rm MSGM}_{4}(\hat \alpha^{t+1}) + \hat \alpha^{t+1}.
\end{equation}

\begin{figure}[t]
    \centering
    \centerline{\includegraphics[width=1.0\linewidth]{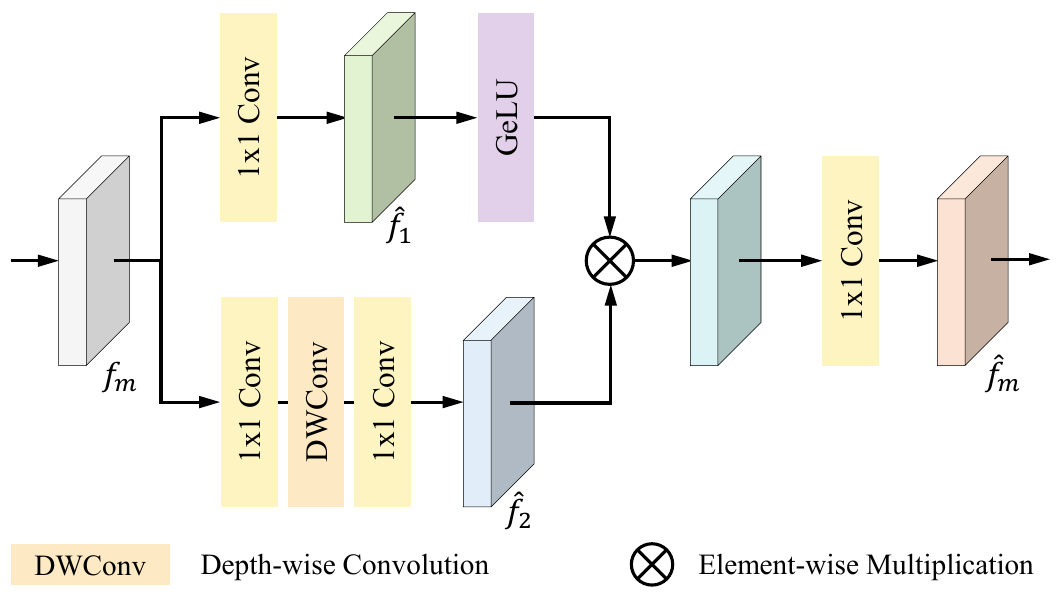}}
    \caption{Details of the proposed MSGM.}
    \label{fig:mgfn}
\end{figure}

\subsection{Mixed-Scale Gating Module}
As shown in Fig.~\ref{fig:mgfn}, the proposed MSGM enhances features by integrating multi-scale representation and a gating mechanism~\cite{zhang2020gated}. 
Given a feature $f_{m} \in R^{C \times H \times W}$, it is processed through parallel branches to capture information at different scales. One branch applies a $1\times1$ convolution, while the other incorporates depth-wise convolution (DWConv) between two $1\times1$ convolutions. 
The output feature $\hat{f}_1$ of the first branch is processed through a GeLU activation function. Then the activated $\hat{f}_1$ is element-wise multiplied with the output feature $\hat{f}_2$ of the second branch to gate $\hat{f}_2$, allowing it to retain relevant information while suppressing irrelevant parts.
The result is then refined through a $1\times1$ convolution:
\begin{equation}
\begin{aligned}
    & \hat{f}_1 = \text{Conv}(f_{m}), \\
    & \hat{f}_2 = \text{Conv}(\text{DWConv}(\text{Conv}(f_{m}))), \\
    & \hat{f}_{m} = \text{Conv}(\sigma(\hat{f}_1) \cdot \hat{f}_2),
\end{aligned}
\end{equation}
where $\sigma$ denotes a GeLU activation function, $\text{Conv}(\cdot)$ denotes a $1\times 1$ convolution and $\cdot$ means element-wise multiplication.

\subsection{Efficient Sparse Attention Module}
Traditional transformers use all features for self-attention calculations, leading to high computational complexity and memory consumption~\cite{zhou2023limits}. 
However, previous works~\cite{NCSR,su2024high} show that only the interconnection between structurally similar patches is beneficial. 
This insight provides a promising opportunity to expand the receptive field while reducing computational complexity, which we efficiently achieve by introducing ESAM.
The input feature $f_{e} \in R^{C \times H \times W}$ is first downsampled using sparse sampling, yielding $X \in R^{C \times H_1 \times W_1}$:
\begin{equation}
    X={\rm Sparse}(f_{e}),
\end{equation}
where ${\rm Sparse(\cdot)}$ operation is implemented using max-pooling, and $H_1\times W_1$ is the spatial dimension of $X$.
The downsampled feature $X$ is then reshaped into $X^{\prime} \in R^{C \times ((\frac{H_1 - M}{M - O} + 1) \times (\frac{W_1 - M}{M - O} + 1)) \times M^2}$ by partitioning $X$ into overlapping $M \times M$ blocks with overlap size $O$. 
Here, $(\frac{H - M}{M - O} + 1) \times (\frac{W - M}{M - O} + 1)$ denotes the total number of blocks. 
Following this, standard self-attention is computed using $X^{\prime}$:
\begin{align}
    Y&= {\rm Softmax}(\frac{Q\cdot K^T}{\sqrt{d_k}})\cdot V,
\end{align}
where $Q={\rm P_Q}(X^{\prime})$, $K={\rm P_K}(X^{\prime})$, and $V={\rm P_V}(X^{\prime})$. 
$P_Q (\cdot)$, $P_K (\cdot)$ and $P_V (\cdot)$ are convolutional layers with the kernel size $1\times1$ and $d_k$ is the dimension of $K$. 
Finally, the result $Y$ is enhanced by a $3\times3$ convolution layer and then upsampled back to its original size:
\begin{equation}
\hat{f}_{e}= \Gamma({\Psi}(Y)),
\end{equation}
where ${\Psi}(\cdot)$ is a $3\times3$ convolution layer and $\Gamma(\cdot)$ denotes bilinear upsampling.

\begin{figure}[t]
    \centering
    \centerline{\includegraphics[width=1.0\linewidth]{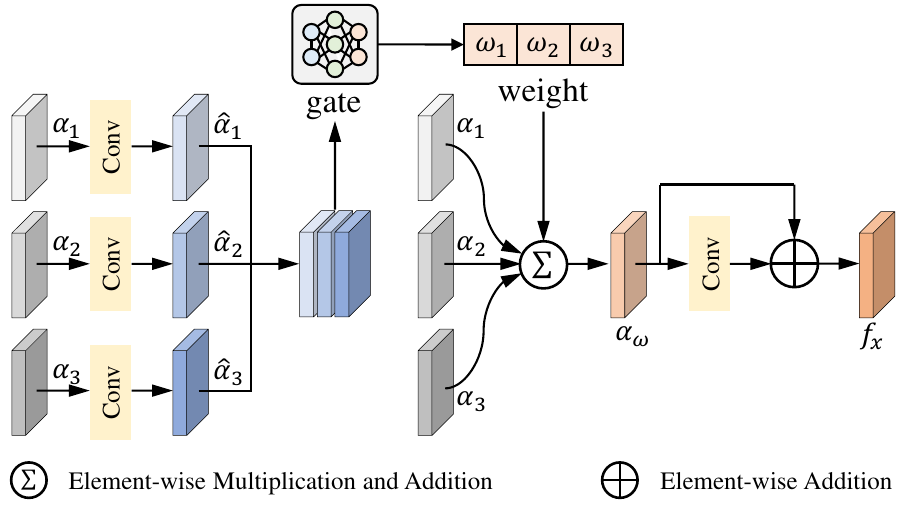}}
    \caption{Details of the proposed MoE-FS. }
    \label{fig:moe}
\end{figure}

\subsection{Mixture-of-Experts based Feature Selector}
The proposed MoE-FS uses a learning-based gating module to integrate features from multiple levels as experts, as illustrated in Fig.~\ref{fig:moe}. 
Using the output $\alpha^{t}$ from the SS-ReM modules as input, MoE-FS combines the distinct characteristics of features from different levels. As mentioned above, we only select 3 features for input. Specifically, we divide the stacked SS-ReM into three groups, corresponding to the encoding block, bottleneck block, and decoding block, and take the output of each block as the final inputs, denoted as $\alpha_1$, $\alpha_2$, and $\alpha_3$.
The features are first processed by a $1\times1$ convolution and fed to the gating module to calculate the feature weights:
\begin{align}
    \hat{\alpha}_{i} &= \text{Conv}_{i}(\alpha_{i}), \quad i=1,2,3,\nonumber\\
    \{{\omega}_{1},{\omega}_{2},{\omega}_{3}\} &= \text{Softmax}(\{\hat{\alpha}_{1},\hat{\alpha}_{2},\hat{\alpha}_{3}\}),
\end{align}
where $\{\cdot\}$ denotes the concatenation operation.
These features are gated based on their corresponding weights to compute the weighted sum $\alpha_{\omega}$. 
It is then passed through a residual block with a $1\times1$ convolution for feature fusion, resulting in the final output feature $f_{\mathbf{x}}$:
\begin{align}
    \alpha_{\omega} &=\alpha_{1}\omega_{1}+\alpha_{2}\omega_{2}+\alpha_{3}\omega_{3},\nonumber\\
    f_{\mathbf{x}} &=\alpha_{\omega}+\text{Conv}(\alpha_{\omega}).
\end{align}

\begin{table*}[t]
\centering
\caption{
Performance comparison of various lightweight SR models on five widely used benchmark datasets.
All PSNR/SSIM values are computed on the Y-channel of the YCbCr color space. 
The FLOPs are measured corresponding to an SR image of size $1280\times 720$ pixels. 
The top three performances are highlighted in red, orange, and yellow backgrounds, respectively.
}
\label{tab:sota_com}
\begin{tabular}{c|c|c|c|c|c|c|c|c|c}
\toprule
\toprule
    & Training &   &  Param.  &  FLOPs  & Set5 & Set14& BSD100 & Urban100 & Manga109 \\
    \multirow{-2}{*}{Method} & Dataset   & \multirow{-2}{*}{Scale} & [K] &  [G]  & PSNR/SSIM     & PSNR/SSIM     & PSNR/SSIM     & PSNR/SSIM     & PSNR/SSIM     \\
\midrule
    CARN~\cite{carn}   & DIV2K &   & 1,592 & 223  & 37.76/0.9590  & 33.52/0.9166  & 32.09/0.8978  & 31.92/0.9256  & 38.36/0.9765 \\    
    EDSR-B~\cite{edsr}   & DIV2K &   & 1,370& 316  & 37.99/0.9604  & 33.57/0.9175  & 32.16/0.8994  & 31.98/0.9272  & 38.54/0.9769  \\
    IMDN~\cite{imdn}   & DIV2K &   & 694  & 159  & 38.00/0.9605  & 33.63/0.9177  & 32.19/0.8996 & 32.17/0.9283  & 38.88/0.9774 \\
    ECBSR-M~\cite{ecbsr}    & DIV2K &   & 596  & 137  & 37.90/0.9615  & 33.34/0.9178  & 32.10/0.9018  & 31.71/0.9250  & -\\
    SMSR~\cite{smsr}   & DIV2K &   & 985  & 132  & 38.00/0.9601  & 33.64/0.9179 & 32.17/0.8990  & 32.19/0.9284 & 38.76/0.9771 \\
    LBNet~\cite{lbnet}   & DIV2K &   & 731  & 153  & 38.05/0.9607  & 33.65/0.9177 & 32.16/0.8994  & 32.30/0.9291 & 38.88/0.9775 \\
    SRFormer~\cite{zhou2023srformer}    & DIV2K &   & 853  & 236  & 38.23/0.9613 & 33.94/0.9209  & 32.36/0.9019  & 32.91/0.9353  & 39.28/0.9785 \\
    CRAFT~\cite{li2023feature}    & DIV2K &   & 737  & 176  & 38.23/0.9615 & 33.92/0.9211  & 32.33/0.9016  & 32.86/0.9343  & 39.39/0.9786 \\
    HiT-SRF~\cite{zhang2024hitsr}    & DIV2K &   & 847  & 227   & 38.26/0.9615 & 34.01/0.9214  & 32.37/0.9023  & 33.13/0.9372  & 39.47/0.9787 \\
    ATD-light~\cite{zhang2024transcending}    & DIV2K &   & 753  & 146   & \cellcolor{yellow!30}38.29/0.9616 & \cellcolor{yellow!30}34.10/0.9217  & \cellcolor{yellow!30}32.39/0.9023  & \cellcolor{orange!30}33.27/0.9375  & \cellcolor{yellow!30}39.52/0.9789 \\
    LAPAR-A~\cite{lapar}& DF2K  &   & 548  & 171  & 38.01/0.9605 & 33.62/0.9183  & 32.19/0.8999 & 32.10/0.9283  & 38.67/0.9772\\
    ShuffleMixer~\cite{shufflemixer}    & DF2K &   & 394  & 91   & 38.01/0.9606 & 33.63/0.9180  & 32.17/0.8995  & 31.89/0.9257  & 38.83/0.9774 \\
    SAFMN~\cite{SAFMN}    & DF2K &   & 228  & 52   & 38.00/0.9605 & 33.54/0.9177  & 32.16/0.8995  & 31.84/0.9256  & 38.71/0.9771 \\
    Omni-SR~\cite{wang2023omni}     & DF2K &   & 772  & 172   & \cellcolor{orange!30}38.29/0.9617 & \cellcolor{red!30}34.27/0.9238  & \cellcolor{orange!30}32.41/0.9026  & \cellcolor{red!30}33.30/0.9386  & \cellcolor{orange!30}39.53/0.9792 \\
    Ours   & DF2K & \multirow{-15}{*}{$\times$2}   & 778 & 164 & \cellcolor{red!30}38.31/0.9632 &   \cellcolor{orange!30}34.20/0.9255 & \cellcolor{red!30}32.43/0.9043 & \cellcolor{yellow!30}33.25/0.9393 & \cellcolor{red!30}39.64/0.9822 \\
\midrule
    CARN~\cite{carn}   & DIV2K &   & 1,592 & 119  & 34.29/0.9255  & 30.29/0.8407  & 29.06/0.8034  & 28.06/0.8493  & 33.50/0.9440 \\
    EDSR-B~\cite{edsr}   & DIV2K &   & 1,555 & 160  & 34.37/0.9270  & 30.28/0.8417  & 29.09/0.8052  & 28.15/0.8527  & 33.45/0.9439  \\
    IMDN~\cite{imdn}   & DIV2K &   & 703  & 72   & 34.36/0.9270  & 30.32/0.8417  & 29.09/0.8046  & 28.17/0.8519  & 33.61/0.9445  \\
    SMSR~\cite{smsr}   & DIV2K &   & 993  & 68   & 34.40/0.9270  & 30.33/0.8412  & 29.10/0.8050  & 28.25/0.8536 & 33.68/0.9445  \\
    LBNet~\cite{lbnet}& DIV2K &   & 736   & 68 & 34.47/0.9277  & 30.38/0.8417  & 29.13/0.8061  & 28.42/0.8559  & 33.82/0.9460  \\
    SRFormer~\cite{zhou2023srformer}    & DIV2K &   & 861  & 105   & 34.67/0.9296 & 30.57/0.8469  & 29.26/0.8099  & 28.81/0.8655  & 34.19/0.9489 \\
    CRAFT~\cite{li2023feature}    & DIV2K &   & 744  & 78   & 34.71/0.9295 & 30.61/0.8469  & 29.24/0.8093  & 28.77/0.8635  & 34.29/0.9491 \\
    HiT-SRF~\cite{zhang2024hitsr}    & DIV2K &   & 855  & 102   & \cellcolor{yellow!30}34.75/0.9300 & 30.61/0.8475  & 29.29/0.8106  & 28.99/0.8687  & 34.53/0.9502 \\
    ATD-light~\cite{zhang2024transcending}    & DIV2K &   & 760  & 65   & 34.74/0.9300 & \cellcolor{yellow!30}30.68/0.8485  & \cellcolor{yellow!30}29.32/0.8109  & \cellcolor{red!30}29.17/0.8709  & \cellcolor{yellow!30}34.60/0.9506 \\
    LAPAR-A~\cite{lapar}& DF2K  &   & 594  & 114  & 34.36/0.9267  & 30.34/0.8421  & 29.11/0.8054  & 28.15/0.8523  & 33.51/0.9441  \\
    ShuffleMixer~\cite{shufflemixer}    & DF2K  &   & 415  & 43   & 34.40/0.9272 & 30.37/0.8423 & 29.12/0.8051 & 28.08/0.8498  & 33.69/0.9448 \\
    SAFMN~\cite{SAFMN}    & DF2K &   & 233  & 23   & 34.34/0.9267 & 30.33/0.8418  & 29.08/0.8048  & 27.95/0.8474  & 33.52/0.9437 \\
    Omni-SR~\cite{wang2023omni}    & DF2K &   & 780  & 78   & \cellcolor{orange!30}34.77/0.9304 & \cellcolor{orange!30}30.70/0.8489  & \cellcolor{orange!30}29.33/0.8111  & \cellcolor{orange!30}29.12/0.8712  & \cellcolor{orange!30}34.64/0.9507 \\
    Ours   & DF2K  & \multirow{-14}{*}{$\times$3}   & 799 & 75 & \cellcolor{red!30}34.79/0.9323 & \cellcolor{red!30}30.71/0.8523 & \cellcolor{red!30}29.35/0.8138 & \cellcolor{yellow!30}29.08/0.8718 & \cellcolor{red!30}34.73/0.9547 \\
\midrule
    CARN~\cite{carn}   & DIV2K &   & 1,592 & 91   & 32.13/0.8937  & 28.60/0.7806  & 27.58/0.7349  & 26.07/0.7837  & 30.47/0.9084 \\
    EDSR-B~\cite{edsr}   & DIV2K &   & 1,518 & 114  & 32.09/0.8938  & 28.58/0.7813  & 27.57/0.7357  & 26.04/0.7849  & 30.35/0.9067  \\
    IMDN~\cite{imdn}   & DIV2K &   & 715  & 41   & 32.21/0.8948  & 28.58/0.7811  & 27.56/0.7353  & 26.04/0.7838  & 30.45/0.9075  \\
    ECBSR-M~\cite{ecbsr}    & DIV2K &   & 603  & 35   & 31.92/0.8946  & 28.34/0.7817  & 27.48/0.7393  & 25.81/0.7773  & -\\
    SMSR~\cite{smsr}   & DIV2K &   & 1006 & 42   & 32.12/0.8932  & 28.55/0.7808  & 27.55/0.7351  & 26.11/0.7868  & 30.54/0.9085  \\
    LBNet~\cite{lbnet}& DIV2K &   & 742   & 39  & 32.29/0.8960  & 28.68/0.7832  & 27.62/0.7382  & 26.27/0.7906  & 30.76/0.9111  \\
    SRFormer~\cite{zhou2023srformer}    & DIV2K &   & 873  & 63   & 32.51/0.8988 & 28.82/0.7872  & 27.73/0.7422  & 26.67/0.8032  & 31.17/0.9165 \\
    CRAFT~\cite{li2023feature}    & DIV2K &   & 753  & 47   & 32.52/0.8989 & 28.85/0.7872  & 27.72/0.7418  & 26.56/0.7995  & 31.18/0.9168 \\
    HiT-SRF~\cite{zhang2024hitsr}    & DIV2K &   & 866  & 58  & 32.55/0.8999 & 28.87/0.7880  & 27.75/0.7432  & 26.80/0.8069  & 31.26/0.9171 \\
    ATD-light~\cite{zhang2024transcending}    & DIV2K &   & 769  & 39   & \cellcolor{orange!30}32.63/0.8998 & \cellcolor{yellow!30}28.89/0.7886  & \cellcolor{yellow!30}27.79/0.7440  & \cellcolor{red!30}26.97/0.8107  & \cellcolor{yellow!30}31.48/0.9198 \\
    LAPAR-A~\cite{lapar}& DF2K  &   & 659  & 94   & 32.15/0.8944  & 28.61/0.7818 & 27.61/0.7366 & 26.14/0.7871 & 30.42/0.9074  \\
    ShuffleMixer~\cite{shufflemixer}    & DF2K  &   & 411  & 28   & 32.21/0.8953 & 28.66/0.7827 & 27.61/0.7366 & 26.08/0.7835  & 30.65/0.9093 \\
    SAFMN~\cite{SAFMN}    & DF2K &   & 240  & 14   & 32.18/0.8948 & 28.60/0.7813  & 27.58/0.7359  & 25.97/0.7809  & 30.43/0.9063 \\
    Omni-SR~\cite{wang2023omni}    & DF2K &   & 792  & 45   & \cellcolor{yellow!30}32.57/0.8993 & \cellcolor{orange!30}28.95/0.7898  & \cellcolor{orange!30}27.81/0.7439  & \cellcolor{orange!30}26.95/0.8105  & \cellcolor{orange!30}31.50/0.9192 \\
    Ours   & DF2K  & \multirow{-15}{*}{$\times$4}   & 794  &  49 & \cellcolor{red!30}32.64/0.9032 & \cellcolor{red!30}28.96/0.7939 & \cellcolor{red!30}27.82/0.7472 & \cellcolor{yellow!30}26.83/0.8098 & \cellcolor{red!30}31.60/0.9242 \\
\bottomrule
\bottomrule
\end{tabular}
\end{table*}

\begin{figure*}[t]
    \centering
    \centerline{\includegraphics[width=1.0\linewidth]{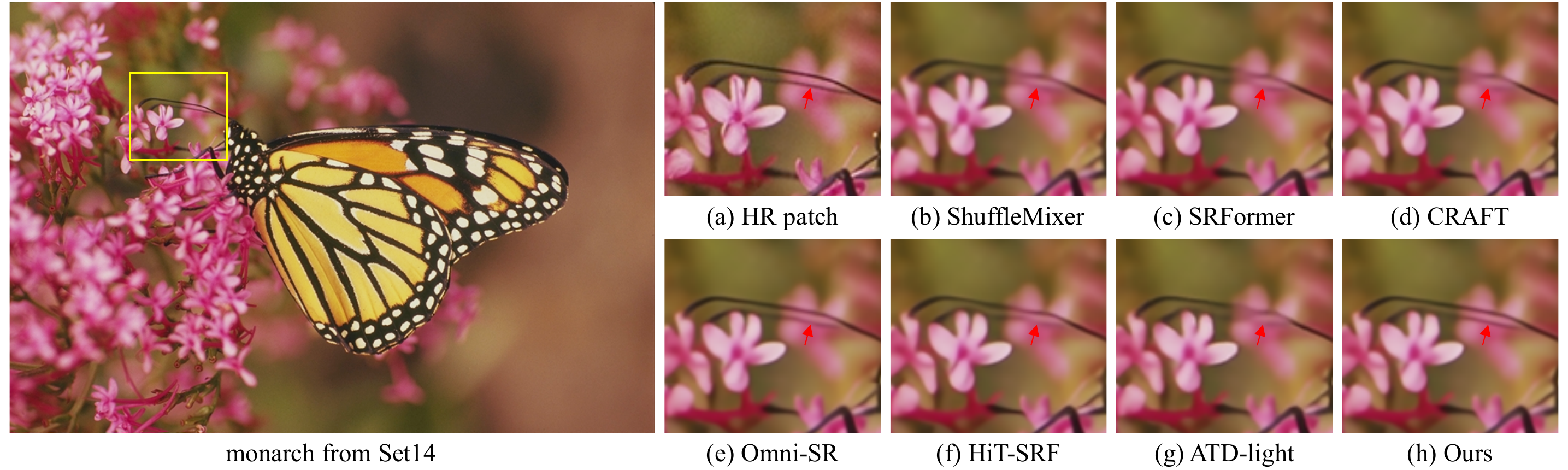}}
    \caption{Visual comparison of $\times 4$ SR on Set14. The proposed SSIU shows clearer structural details.}
    \label{fig:set14}
\end{figure*}

\begin{figure*}[t]
    \centering
    \centerline{\includegraphics[width=1.0\linewidth]{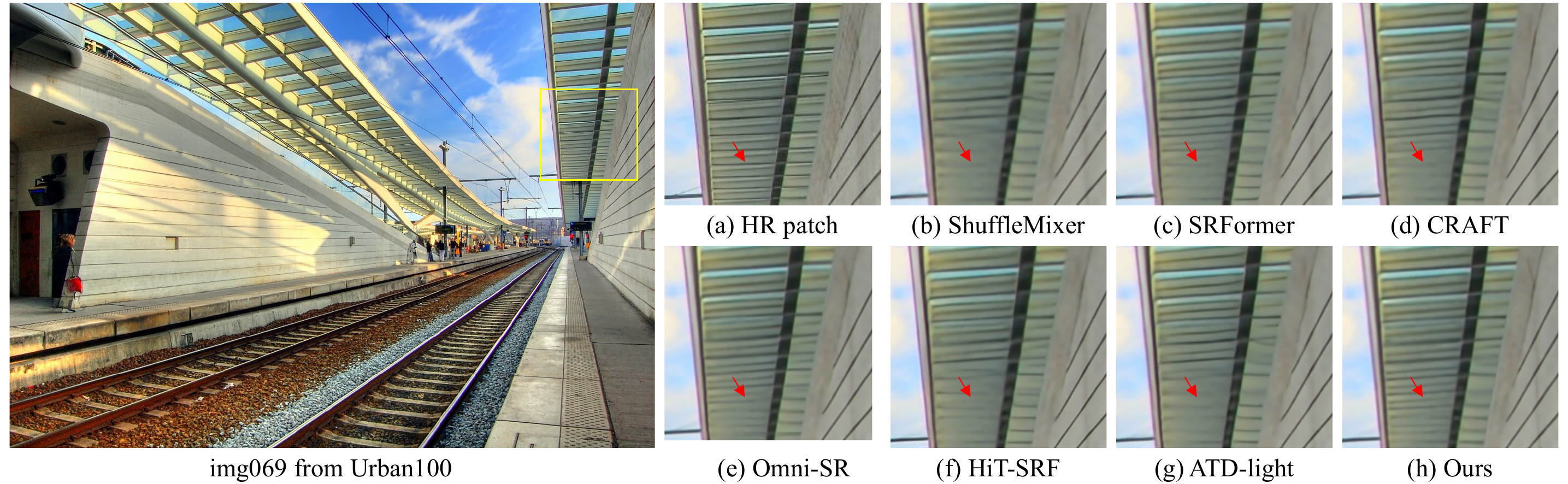}}
    \caption{Visual comparison of $\times 4$ SR on Urban100. The proposed SSIU shows clearer structural details.}
    \label{fig:urban100}
\end{figure*}

\subsection{Training Objectives}
The proposed SSIU is optimized using the objective:
\begin{equation}
    \label{equ:loss}
    L = L_{\text{1}} + \lambda L_{\text{f}},
\end{equation}
where $L_{\text{1}}$ measures the $L_1$ distance between the SR result and the ground truth, and $L_{\text{f}}$ computes the $L_1$ distance between their Fast Fourier Transforms (FFT)-transformed versions.
The constant $\lambda$ balances the relative importance of $L_{\text{f}}$.

\begin{figure*}[t]
    \centering
    \centerline{\includegraphics[width=1.0\linewidth]{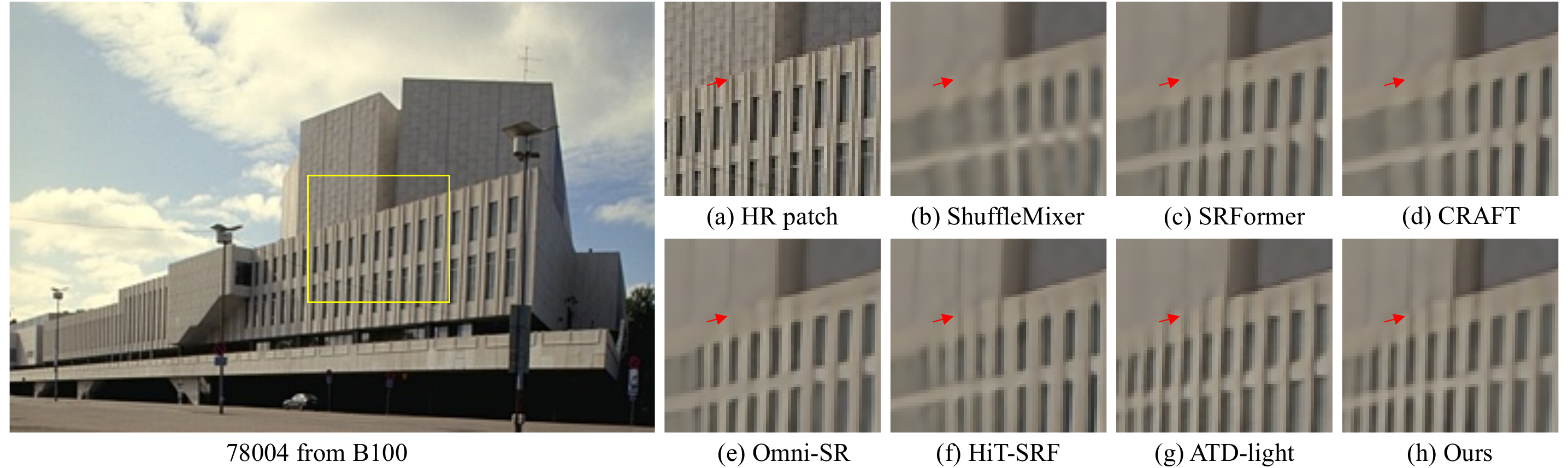}}
    \caption{Visual comparison of $\times 4$ SR on B100. The proposed SSIU shows clearer structural details.}
    \label{fig:B100}
\end{figure*}

\begin{figure*}[t]
    \centering
    \centerline{\includegraphics[width=1.0\linewidth]{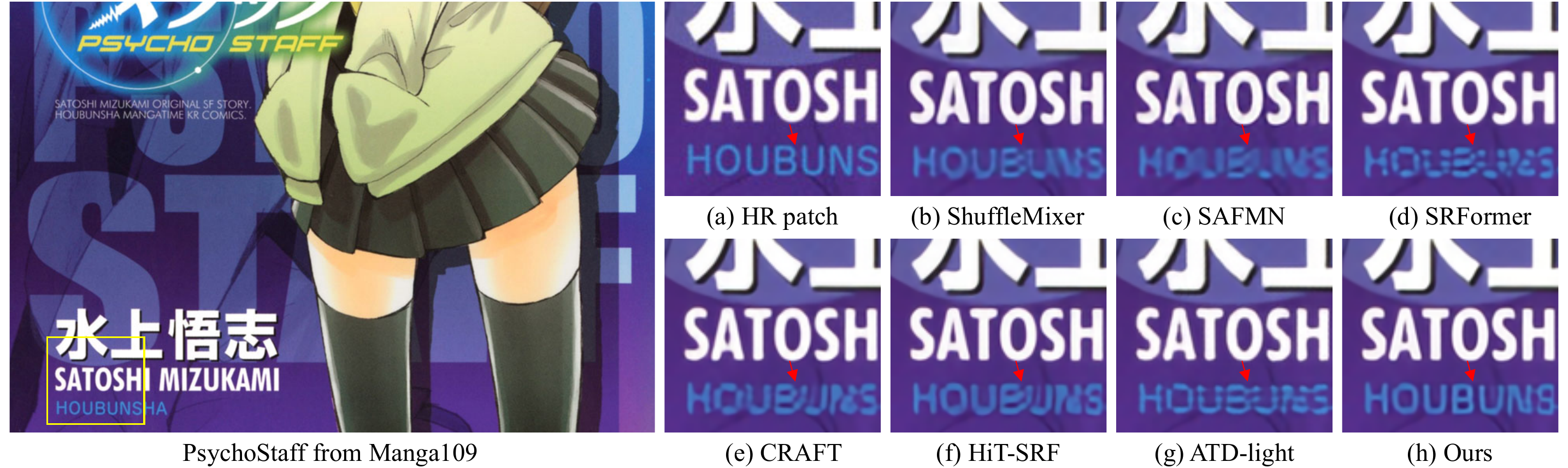}}
    \caption{Visual comparison of $\times 4$ SR on Manga109. The proposed SSIU shows clearer structural details.}
    \label{fig:manga109}
\end{figure*}

\section{Experiment}
\label{sec5:results}

\subsection{Experimental Setup}

\subsubsection{Dataset}
Following previous works~\cite{shufflemixer, lapar}, we train our model on the DF2K dataset, which combines DIV2K~\cite{dataset_div2k} and Flickr2K~\cite{edsr}, resulting in a total of 3,450 high-quality images (800 from DIV2K and 2,650 from Flickr2K).
The LR images are generated by applying bicubic downscaling to the corresponding HR images. 
To thoroughly assess our method’s performance, we perform evaluations on five widely-used benchmark datasets: Set5~\cite{dataset_set5}, Set14~\cite{dataset_set14}, BSD100~\cite{dataset_b100}, Urban100~\cite{dataset_urban100}, and Manga109~\cite{dataset_manga109}.

\subsubsection{Implementation Details}
In the proposed SSIU, we employ SS-ReM modules with 64 feature channels. 
For training, 40 patches of size $64 \times 64$ are randomly cropped from LR images and paired with corresponding HR patches. 
These training patch pairs are augmented by random horizontal flipping and rotation. The proposed model is trained using the Adam optimizer~\cite{adam}.
The learning rate is initialized at 1$e$-3 and decays to 1$e$-6 following the Cosine Annealing schedule~\cite{loshchilov2016sgdr}. 
The $\lambda$ is set to 0.01 in all of our experiments.
All experiments are conducted with the PyTorch framework on an NVIDIA GeForce RTX 4090 GPU.

\subsubsection{Benchmarks}
To comprehensively evaluate the performance of the proposed SSIU, we compare it with classic and state-of-the-art lightweight SR models, including CARN~\cite{carn}, EDSR-baseline~\cite{edsr}, IMDN~\cite{imdn}, ECBSR~\cite{ecbsr}, SMSR~\cite{smsr}, LBNet~\cite{lbnet}, SRFormer~\cite{zhou2023srformer}, CRAFT~\cite{li2023feature}, HiT-SRF~\cite{zhang2024hitsr}, ATD-light~\cite{zhang2024transcending}, LAPAR~\cite{lapar}, ShuffleMixer~\cite{shufflemixer}, SAFMN~\cite{SAFMN}, and Omni-SR~\cite{wang2023omni}.

\subsubsection{Evaluation Metrics}
To assess the quality of the SR images, we employ peak signal-to-noise ratio (PSNR) and structural similarity index (SSIM) metrics. 
Following the setting of existing works, all PSNR and SSIM values are computed based on the luminance (Y) channel of the images in the YCbCr color space.
PSNR evaluates the quality in terms of fidelity, while SSIM focuses on perceptual quality.

\subsection{Comparisons with State-of-the-Art Methods}
\subsubsection{Quantitative Comparison}
Table~\ref{tab:sota_com} presents quantitative comparisons of various SR models for upscaling factors $\times 2$, $\times 3$, and $\times 4$ on five widely used benchmark datasets. 
It provides details on the number of model parameters, floating-point operations (FLOPs), and training datasets for each model. 
The FLOPs are calculated based on the SR of an image of $1280\times 720$ pixels. 
The proposed SSIU model, leveraging structural similarity-inspired SR optimization and the efficient SS-ReM, demonstrates superior performance with reduced parameters and lower computational complexity. 
Several critical conclusions can be drawn from Table~\ref{tab:sota_com}. 
Firstly, compared to classic and state-of-the-art lightweight SR algorithms, SSIU achieves better performance with similar parameters and FLOPs. 
Particularly, for $\times 4$ SR on the Manga109 dataset, the performance gain of the proposed SSIU over LBNet, CRAFT, and ATD-light is 0.76dB, 0.25dB, and 0.12dB, respectively. 
Secondly, for the $\times 2$ upscaling factor, SSIU performs competitively against SRFormer, Omni-SR, and HiT-SRF, while requiring fewer parameters or FLOPs. 
Although models like ShuffleMixer and SAFMN have fewer parameters, SSIU significantly outperforms them on all benchmark datasets.
Finally, for $\times 3$ and $\times 4$ upscaling, SSIU consistently achieves superior or competitive performance across all datasets.
Overall, the results in Table~\ref{tab:sota_com} demonstrate that SSIU offers a compelling trade-off between efficiency and reconstruction quality compared to existing state-of-the-art methods.

\begin{figure}[ht]
    \centering
    \centerline{\includegraphics[width=1.0\linewidth]{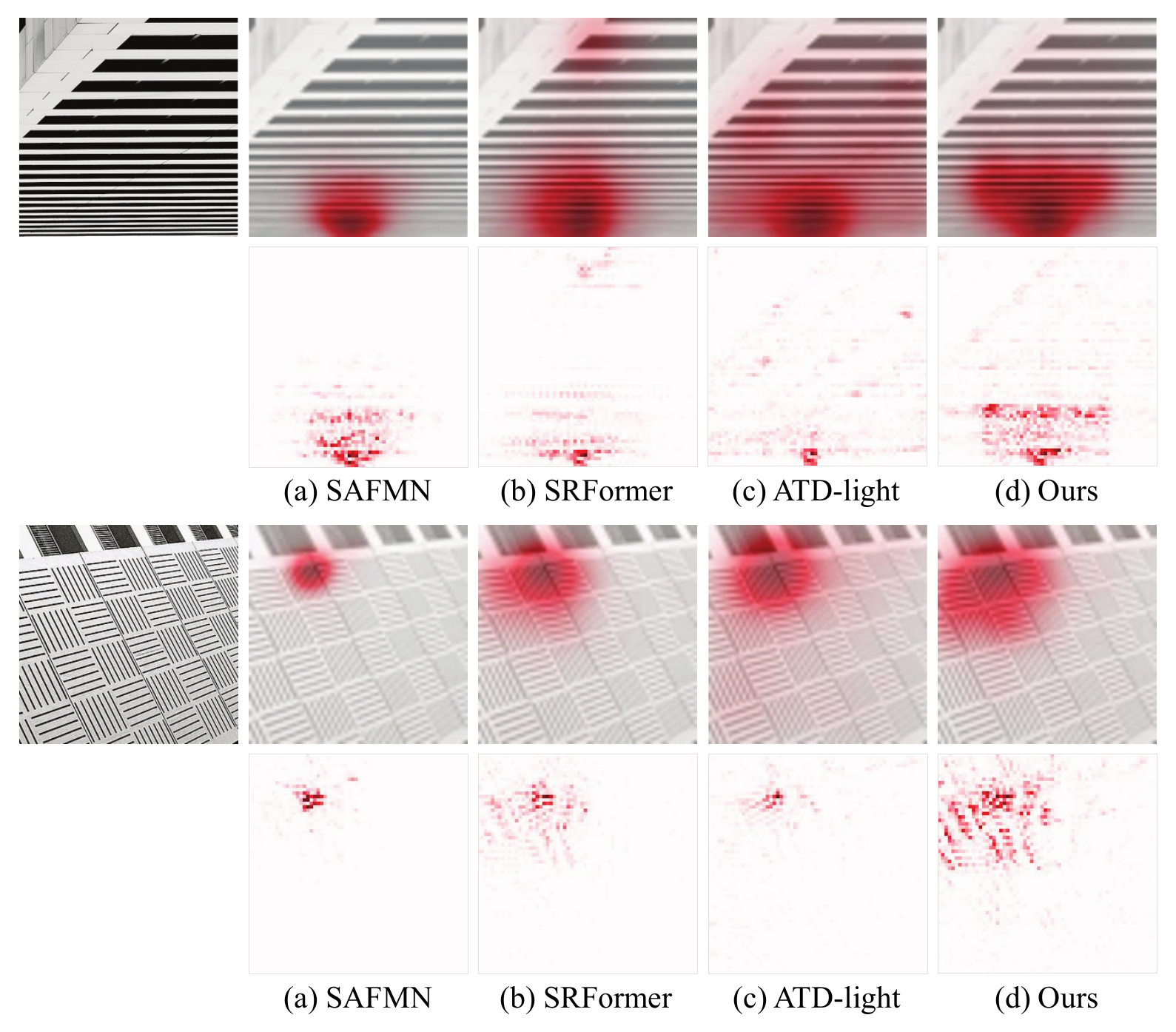}}
    \caption{Visual comparison of local attribution maps (LAM) of multiple methods in the same target area.}
    \label{fig:heatmap}
\end{figure}

\subsubsection{Qualitative Comparison}
The visual comparison results of the ×4 upscaling factor on the Set14, Urban100, B100, and Manga109 datasets are presented in Fig.~\ref{fig:set14}, Fig.~\ref{fig:urban100}, Fig.~\ref{fig:B100}, and Fig.~\ref{fig:manga109}, with the zoomed-in detail results.
The qualitative comparison highlights that our proposed SSIU consistently produces sharper edges, enhanced structures, and textures.
Specifically, from Fig.~\ref{fig:set14} to Fig.~\ref{fig:manga109}, we observe that SSIU generates more accurate details compared to other methods, particularly in lines and structures.   
This highlights SSIU's superior performance in preserving and enhancing crucial features in upscaled images, further validating the effectiveness of the structural similarity-inspired unfolding framework, which enables our model to capture and refine complex structural details in SR images.

\begin{table}[t]
\centering
\caption{PSNR (dB), Memory Usage (M), and Inference Time (ms) Comparison for $\times$4 SR. $\text{GPU Mem.}$ is the peak GPU memory consumption during inference, while $\text{Avg. Time}$ is the average processing time for 100 SR images of size 480$\times$320 pixels.}
\label{tab:inference_time}
\tabcolsep2pt
\begin{tabular}{ccccccc}
\toprule
\toprule
    Metrics & SRFormer & CRAFT & Omni-SR & HiT-SRT & ATD-light & SSIU(Ours) \\
\midrule
    PSNR & 27.73 & 27.72 & \cellcolor{orange!30}27.81 & 27.75 & \cellcolor{yellow!30}27.79 & \cellcolor{red!30}27.82\\
    GPU Mem. &  4181 & \cellcolor{yellow!30}1586 & 1597 & 1878 & \cellcolor{orange!30}753 & \cellcolor{red!30}224 \\
    Avg. Time  & 42.58 & \cellcolor{yellow!30}33.94 & \cellcolor{orange!30}19.61 & 47.96 & 41.97 & \cellcolor{red!30}13.83 \\
\bottomrule
\bottomrule
\end{tabular}
\end{table}

\begin{table*}[t]
\centering
\small
\caption{Performance comparison with and without the MOE-FS and ESAM. SAM denotes the self-attention module.}
\label{tab:module}
\tabcolsep4pt
\begin{tabular}{ccc|c|c|cc|cc|cc|cc|cc}
\toprule
\toprule
    \multicolumn{1}{c}{\multirow{2}{*}{ESAM}} & \multicolumn{1}{c}{\multirow{2}{*}{SAM}} & \multicolumn{1}{c}{\multirow{2}{*}{MOE-FS}} & \multicolumn{1}{|c}{\multirow{2}{*}{Flops}} & \multicolumn{1}{c|}{\multirow{2}{*}{Param.}} & \multicolumn{2}{c}{Set5} & \multicolumn{2}{c}{Set14} & \multicolumn{2}{c}{B100} & \multicolumn{2}{c}{Urban100} & \multicolumn{2}{c}{Manga109} \\
    \multicolumn{1}{c}{} & \multicolumn{1}{c}{} & \multicolumn{1}{c}{} & \multicolumn{1}{|c}{}  & \multicolumn{1}{c|}{}  & PSNR & SSIM & PSNR & SSIM & PSNR & SSIM & PSNR  & SSIM & PSNR & SSIM          \\
\midrule
    \checkmark &  &  & \cellcolor{red!30}48 & \cellcolor{red!30}777.83 & \cellcolor{red!30}32.42  & \cellcolor{yellow!30}0.9012 & \cellcolor{yellow!30}28.88 & \cellcolor{yellow!30}0.7818  & \cellcolor{yellow!30}27.76 & \cellcolor{yellow!30}0.7452 & \cellcolor{yellow!30}26.63 & \cellcolor{yellow!30}0.8029  & \cellcolor{yellow!30}31.29  & \cellcolor{yellow!30}0.9210   \\
    \checkmark &  & \checkmark & \cellcolor{orange!30}49 & \cellcolor{yellow!30}794.47 & \cellcolor{red!30}32.64  & \cellcolor{red!30}0.9032 & \cellcolor{orange!30}28.96 & \cellcolor{orange!30}0.7939  & \cellcolor{orange!30}27.82 & \cellcolor{orange!30}0.7472 & \cellcolor{orange!30}26.83 &\cellcolor{orange!30}0.8098  & \cellcolor{orange!30}31.60  & \cellcolor{orange!30}0.9242     \\
     & \checkmark & \checkmark & \cellcolor{yellow!30}54 & \cellcolor{orange!30}788.71 & \cellcolor{orange!30}32.62  & \cellcolor{orange!30}0.9031 & \cellcolor{red!30}28.99 & \cellcolor{red!30}0.7944  & \cellcolor{red!30}27.83 & \cellcolor{red!30}0.7473 & \cellcolor{red!30}26.90 & \cellcolor{red!30}0.8106  & \cellcolor{red!30}31.64  & \cellcolor{red!30}0.9247    \\
\bottomrule
\bottomrule
\end{tabular}
\end{table*}

\subsubsection{Inferencce Time}
To demonstrate the computational efficiency of our method, we compare the peak GPU memory consumption and the average inference time with several state-of-the-art approaches that have comparable parameter sizes. 
Specifically, the maximum GPU memory usage refers to the highest memory consumed during inference, and the average inference time is computed over 100 SR images of resolution $480 \times 320$ pixels, evaluated on an NVIDIA GeForce RTX 4090 GPU. 
As shown in Table~\ref{tab:inference_time}, our method achieves the lowest average inference time and the least GPU memory consumption among all compared models. 
These results confirm that our approach introduces minimal computational overhead while delivering superior performance, underscoring the effectiveness of the proposed SSIU framework in reducing model complexity without compromising quality.

\subsection{Ablation Studies and Discussions}
To better understand the workings of SSIU, we conduct comprehensive ablation studies to evaluate the roles of its various components, discussing their functionalities, contributions, and key parameter choices.

\subsubsection{Effectiveness of the MoE-FS} 
To validate the effectiveness of the proposed MoE-FS module, we perform ablation studies, as presented in the first and second rows of Table~\ref{tab:module}.  
Although integrating MoE-FS introduces a slight increase in model parameters and computational complexity, it yields significant performance improvements across all benchmark datasets.  
These results demonstrate that aggregating multi-level features through the MoE-FS mechanism is beneficial for enhancing SR performance.

\subsubsection{Effectiveness of the ESAM} 
In this experiment, we investigate the impact of the sparsity operation in the proposed ESAM on performance, comparing it with the traditional self-attention module (SAM) that does not incorporate sparsity. 
From the results presented in the second and third rows of Table~\ref{tab:module}, we draw the following conclusions: 
Firstly, the sparsity operation in ESAM significantly reduces the computed feature quantity by eliminating redundant information while retaining the most informative features, reducing Flops by 10.2\% compared to SAM.  
Secondly, compared with the traditional self-attention module, the proposed ESAM maintains the performance on multiple benchmarks with a slight decrease. 
This shows that only long-distance information with a similar structure is needed to reconstruct the current pixel information, which effectively proves the rationality of our method. 
To analyze the information aggregation capability of ESAM, we utilize local attribution maps (LAM)~\cite{gu2021interpreting}, a visualization tool that identifies pixels strongly influencing SR results, also known as information regions. 
Larger information regions indicate better aggregation ability. 
As shown in Fig.~\ref{fig:heatmap}, we compare the information regions of various methods within the same target area. 
Our method demonstrates a larger information region than the others, highlighting that ESAM is more effective at modeling long-distance context using the attention mechanism.

\begin{table*}[t]
\centering
\small
\caption{Performance comparison of variants of different loss functions.}
\label{tab:loss_ablation}
\begin{tabular}{ccc|cc|cc|cc|cc|cc}
\toprule
\toprule
    \multicolumn{1}{c}{\multirow{2}{*}{loss}} & \multicolumn{1}{c}{\multirow{2}{*}{Flops}} & \multicolumn{1}{c|}{\multirow{2}{*}{Param.}} & \multicolumn{2}{c}{Set5} & \multicolumn{2}{c}{Set14} & \multicolumn{2}{c}{B100} & \multicolumn{2}{c}{Urban100} & \multicolumn{2}{c}{Manga109} \\
    \multicolumn{1}{c}{} & \multicolumn{1}{c}{} & \multicolumn{1}{c|}{} & PSNR       & SSIM        & PSNR        & SSIM        & PSNR       & SSIM        & PSNR         & SSIM          & PSNR         & SSIM          \\
\midrule
    $L_{\text{1}}$      & 49   & 794.47    & \cellcolor{orange!30}32.56      & \cellcolor{orange!30}0.9022      & \cellcolor{orange!30}28.87       & \cellcolor{orange!30}0.7928      & \cellcolor{orange!30}27.79     & \cellcolor{orange!30}0.7468      & \cellcolor{orange!30}26.81        & \cellcolor{orange!30}0.8096        & \cellcolor{orange!30}31.44        & \cellcolor{orange!30}0.9234        \\
    $L_{\text{1}}$+$\lambda L_{\text{f}}$   & 49   & 794.47    & \cellcolor{red!30}32.64      & \cellcolor{red!30}0.9032      & \cellcolor{red!30}28.96       & \cellcolor{red!30}0.7939      & \cellcolor{red!30}27.82      & \cellcolor{red!30}0.7472      & \cellcolor{red!30}26.83        & \cellcolor{red!30}0.8098        & \cellcolor{red!30}31.60        & \cellcolor{red!30}0.9242       \\
\bottomrule
\bottomrule
\end{tabular}
\end{table*}

\subsubsection{Effectiveness of the FFT Loss} Previous studies usually optimize neural networks using the $L_1$ distance. However, in this work, we explore the use of L1 distance in the frequency domain to encourage the network to focus more on high frequencies. From the results presented in Table~\ref{tab:loss_ablation}, it can be observed that the FFT loss contributes to the improved performance of our network.

\section{Conclusion}   
\label{sec:conclusion}
Deep learning methods in SR often focus on enlarging the receptive fields to enhance performance, which can lead to increased model complexity. 
Our work seeks to develop a more effective method by activating pixels more sparsely. 
We employ an optimization function inspired by nonlocal structural similarity to build a lightweight SR network. 
By adopting an unfolding paradigm, our model establishes a step-by-step iterative process. 
In this process, mixed-scale gating modules and a sparse attention module are applied together; the former imposes sparsity and structural similarity constraints, while the latter achieves sparse long-range modeling. 
Additionally, we introduce a mixture-of-experts-based feature selector to gate feature information across different scales. 
Extensive experiments demonstrate that our unfolding-inspired network achieves both effectiveness and efficiency, outperforming current state-of-the-art models in terms of performance and inference time.

\bibliography{main} 
\bibliographystyle{IEEEtran}

\end{document}